\pdfoutput=1
\documentclass[conference]{IEEEtran}

\usepackage{multirow}
\usepackage{amsmath}
\usepackage{wasysym}

\ifCLASSINFOpdf
  \usepackage[pdftex]{graphicx}
  \graphicspath{{figures/}}
  \DeclareGraphicsExtensions{.pdf,.jpeg,.png}
\else
  \usepackage[dvips]{graphicx}
  \graphicspath{{figures/}}
  \DeclareGraphicsExtensions{.eps}
\fi

\usepackage{color,xcolor}

\usepackage{cases}

\usepackage{algorithm}
\usepackage{algpseudocode}

\usepackage{threeparttable}
\usepackage{array}

\ifCLASSOPTIONcompsoc
  \usepackage[caption=false,font=normalsize,labelfont=sf,textfont=sf]{subfig}
\else
  \usepackage[caption=false,font=footnotesize]{subfig}
\fi

\ifCLASSOPTIONcaptionsoff
 \usepackage[nomarkers]{endfloat}
 \let\MYoriglatexcaption\caption
 \renewcommand{\caption}[2][\relax]{\MYoriglatexcaption[#2]{#2}}
\fi

\usepackage{url}

\begin{document}
\title{Radio Frequency Fingerprint Identification Based on Denoising Autoencoders}
\author{\IEEEauthorblockN{Jiabao~Yu\IEEEauthorrefmark{1}\IEEEauthorrefmark{3}\IEEEauthorrefmark{4},
Aiqun Hu\IEEEauthorrefmark{2}, Fen Zhou\IEEEauthorrefmark{4}\IEEEauthorrefmark{3},
Yuexiu Xing\IEEEauthorrefmark{1}, Yi Yu\IEEEauthorrefmark{4}\IEEEauthorrefmark{5},
Guyue Li\IEEEauthorrefmark{2}, Linning Peng\IEEEauthorrefmark{2}}\IEEEauthorblockA{\IEEEauthorrefmark{1}School of Information Science and Engineering, 
Southeast University, 210096 Nanjing, China.\\
\IEEEauthorrefmark{2}School of Cyber Science and Engineering, Southeast
University, 210096 Nanjing, China.\\
\IEEEauthorrefmark{3}CERI-LIA, Universit\'e d'Avignon, 84000 Avignon,
France\\
\IEEEauthorrefmark{4}Institut Sup\'erieur d'Electronique de Paris,
75006 Paris, France \\
\IEEEauthorrefmark{5}Conservatoire National des Arts et M\'etiers,
75003 Paris, France \\
Email: \{yujiabao, aqhu, yxxing, guyuelee, pengln\}@seu.edu.cn, fen.zhou@\{isep,
univ-avignon\}.fr, yi.yu@isep.fr
}}
\maketitle
\begin{abstract}
Radio Frequency Fingerprinting (RFF) is one of the promising passive
authentication approaches for improving the security of the Internet
of Things (IoT). However, with the proliferation of low-power IoT
devices, it becomes imperative to improve the identification accuracy
at low SNR scenarios. To address this problem, this paper proposes
a general Denoising AutoEncoder (DAE)-based model for deep learning
RFF techniques. Besides, a partially stacking method is designed to
appropriately combine the semi-steady and steady-state RFFs of ZigBee
devices. The proposed Partially Stacking-based Convolutional DAE (PSCDAE)
aims at reconstructing a high-SNR signal as well as device identification.
Experimental results demonstrate that compared to Convolutional Neural
Network (CNN), PSCDAE can improve the identification accuracy by 14\%
to 23.5\% at low SNRs (from -10 dB to 5 dB) under Additive White Gaussian
Noise (AWGN) corrupted channels. Even at SNR = 10 dB, the identification
accuracy is as high as 97.5\%.
\end{abstract}


\IEEEpeerreviewmaketitle{}

\section{Introduction\label{sec:Introduction}}

Nowadays, the Internet of things (IoT) is becoming more and more ubiquitous
in our everyday lives. IoT integrates physical devices into the core
networks to allow them to communicate with each other and offer efficient
services. However, with the deployment of IoT networks, various security
risks have emerged as IoT devices can provide and/or have access to
private information. One of the profound security challenges in IoT
networks is identity authentication, which is the first line of defense
against intruders. Identity vulnerabilities in the IoT networks contain
these main facets as follows. First, cryptographic keys or certificates
cannot be efficiently distributed due to the limited computing power
leading to weak authentication \cite{Li-2019-Physical}. Second, default
keys or credentials can be brute-forced by high-computing-power attackers,
extracted from device firmware or mobile apps, or intercepted at login
\cite{Bastos-2018-Internet}. Third, the ID number such as MAC address
in the header needs extra spectral or power resources, which is limited
in IoT applications, to be transmitted \cite{Morin-2019-Transmitter}.
Hence, a passive authentication mechanism without cryptographic materials
or IDs may be the future of IoT identity authentication.

Radio Frequency Fingerprinting (RFF) is a promising passive authentication
technique to identify transmitters by extracting device-specific features/fingerprints
from Radio Frequency (RF) signals. RFF is the inherent attribute of
the device's hardware variability in the RF frontend \cite{Yu-2019-A}.
It is unique for each transmitter and is arduous to impersonate. A
lot of research has been executed in this area since the 1990s. Traditional
RFF methods can be grouped into transient approaches and steady-state
approaches based on their target signal region. More recently, some
studies have embraced Deep Learning (DL) for RFF identification. These
existing DL RFF techniques all belong to steady-state approaches.
In these approaches, minimal preprocessing is carried out on the down-converted
baseband signals and then sent to Neural Networks (NNs) for feature
extraction and classification \cite{Hanna-2019-Deep}. Nevertheless,
depending on the input type of NN, DL RFF techniques can be further
categorized into Time-series-based DL RFFs (TDL RFFs) and Image-based
DL RFFs (IDL RFFs).

TDL RFFs always adopt baseband In-phase/Quadrature (I/Q) samples,
which are concatenated in one channel or separate in I/Q channels,
as the network input. In \cite{Robyns-2017-Physical-layer}, MultiLayer
Perceptron (MLP) and Convolutional Neural Network (CNN) were applied
to each symbol alone to discern 22 LoRa devices. Subsequently, Merchant
\emph{et al.} input the time-domain complex baseband error signal
to a CNN to identify seven ZigBee devices against primary user emulation
attacks in cognitive radio networks \cite{Merchant-2018-Deep}. Long
Short Term Memory (LSTM) was also used to learn higher-order correlations
between the signal samples to identify USRPs \cite{Wu-2018-Deep}.
In \cite{Yu-2019-A}, a multi-sampling CNN was proposed to extract
multi-scale features from the well-synchronized preambles for discerning
54 ZigBee devices.

In IDL RFFs, time series are further transformed to images based on
various techniques before feeding into networks. To name only a few,
Recurrence Plot (RP), Continuous Wavelet Transform (CWT), Short-Time
Fourier Transform (STFT), and Hilbert Transform (HT) have been employed
to generate images. After transforming, these images were sent to
a deep network such as CNN, Deep Neural Network (DNN), Deep Residual
Network (DRN), or Multi-Stage Training (MST) for feature extraction
and transmitter identification \cite{Baldini-2019-Comparison,Baldini-2019-Physical,Youssef-2018-Machine,Pan-2019-Specific}.

Although various DL RFFs have been demonstrated to achieve excellent
identification in high SNR regions, their performance in low SNRs
is far from satisfactory. For instance, at SNR = 10 dB, the classification
accuracies were only 73.73\% for seven ZigBee devices \cite{Merchant-2018-Deep},
38\% for 12 mobile phones \cite{Baldini-2019-Comparison}, 58\% for
seven USRPs \cite{Huang-2018-Deep}, and 65\% for five simulated radio
emitters \cite{Pan-2019-Specific}, respectively. However, since most
of the IoT devices are battery-powered, the transmitting power is
relatively low, they usually need to work in low SNR scenarios. Therefore,
it has a practical significance to study DL RFFs in low SNR scenarios.

To overcome these shortcomings, this paper proposes a general DL model
based on Denoising AutoEncoder (DAE) and a dedicated partially stacking
method for ZigBee devices. First, we perform synchronization and compensation
on the received ZigBee baseband signals. Then the steady-state symbols
in the preamble are stacked and concatenated to the semi-steady preamble
symbols. Thereafter, we propose a Convolutional Denoising AutoEncoder
(CDAE)-based network to extract fingerprints from these concatenated
symbols for identifying 27 Ti CC2530 ZigBee devices.

The main contributions are summarized as follows:
\begin{itemize}
\item We propose a universal DAE-based architecture for all the DL RFF techniques
to enhance their performance, especially at low and medium SNRs. Our
DAE-based approach can be trained by jointly minimizing the reconstruction
error and the classification loss.
\item Inspired by stacking spread sequences to enhance SNR \cite{Xing-2018-On},
we partially stack the steady-state symbols in the ZigBee preamble
rather than stack all the preamble symbols. Partially stacking and
concatenation are efficient for combining the semi-steady RFF and
the steady-state RFF.
\item We further investigate the performance of our proposed Partially Stacking-based
CDAE (PSCDAE) approach by selecting two convolutional layers as the
encoder of CDAE and a dense layer with Softmax as the classifier.
Simulation results demonstrate that the proposed PSCDAE outperforms
the original CNN method under all training scenarios. When identifying
27 ZigBee devices, it achieves an accuracy of 97.5\% even at SNR =
10 dB.
\end{itemize}
\indent The remainder of this paper is organized as follows. Section
II introduces the novel DAE-based DL RFF approach. Section III discusses
the partially stacking method for ZigBee devices. Section IV shows
the performance evaluation of the proposed scheme. Finally, Section
V concludes this paper.

\section{RFF Based on Denoising Autoencoders}

\begin{figure}[htbp]
\centering\includegraphics[width=3in]{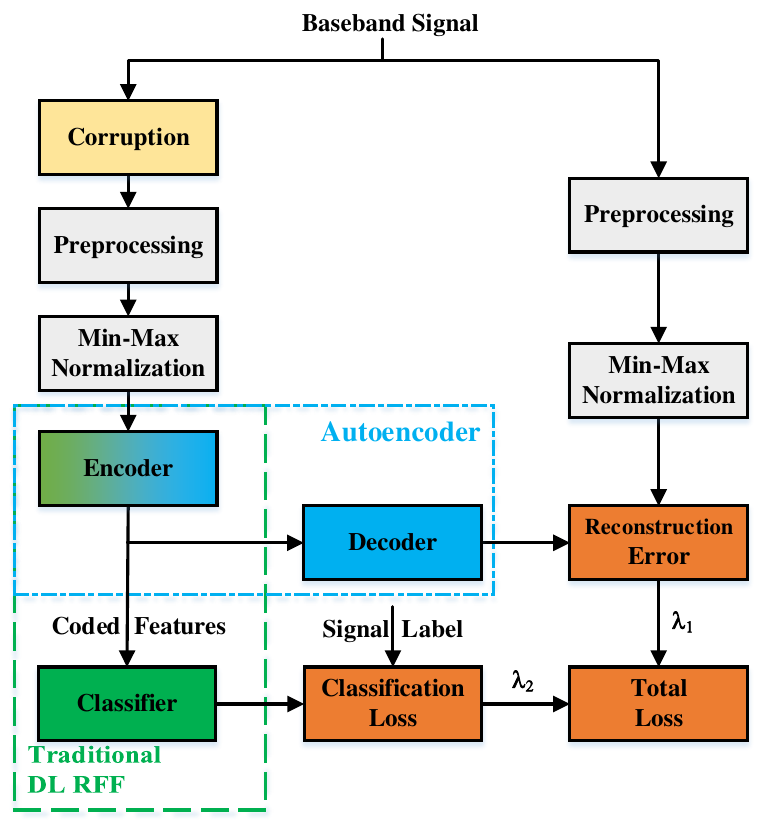}
\caption{Denoising AutoEncoder-based DL RFF architecture.}
\label{DAE-DL-RFF}
\end{figure}

An AutoEncoder (AE) is a neural network designed to learn high-level
data representation by the encoder and reconstruct the input from
that representation by the decoder in an unsupervised manner \cite{Bengio-2007-Greedy}.
By introducing input corruption such as masking noise, effective features
relatively stable and robust to the corruption can be extracted by
DAE \cite{Vincent-2008-Extracting}. However, AE is generally used
for pretraining and initialization of classification networks. Here,
we propose a novel DAE-based DL RFF approach, in which the encoder,
decoder, and classifier are trained simultaneously, as shown in Fig.
\ref{DAE-DL-RFF}. It is worth noting that a conventional DL RFF method
only includes an encoder for feature extraction and a classifier for
identification without a decoder for reconstruction.

Assume the received baseband signal $\mathbf{s}$ in the training
is ideal, which means there is no external wireless channel. It can
be denoted as
\begin{equation}
\mathbf{s}=RFF(\mathbf{x}),
\end{equation}
where the mathematical function $RFF(\cdotp)$ represents the device
fingerprint, and $\mathbf{x}$ is the transmitted data. Then we will
first corrupt the input $\mathbf{s}$ to simulate the channel influence
as
\begin{equation}
\tilde{\mathbf{s}}=H(\mathbf{s}),
\end{equation}
where $H(\cdotp)$ can be an Additive White Gaussian Noise (AWGN)
channel or a multipath channel. Thereafter, various preprocessing
approaches including synchronization, frequency and phase offset compensation
\cite{Peng-2018-Design}, and transforms (CWT, STFT, HT, etc.) can
be applied to the corrupted signal:
\begin{equation}
\mathbf{s'}=Pre(\tilde{\mathbf{s}}),
\end{equation}
where $Pre(\cdotp)$ denotes the preprocessing. Its output $\mathbf{s'}$
can be either a time series or an image, thus our method fits for
TDL RFFs and IDL RFFs. Additionally, a min-max normalization is indispensable,
since the output of the decoder introduced later is always between
0 and 1. The minimum and maximum values in each dimension of the preprocessed
vector $\mathbf{s'}$ in the training dataset are used and saved for
the min-max normalization:
\begin{equation}
\begin{split}\mathbf{z}= & \:min\_max\_norm(\mathbf{s^{'}})\\
= & \:\frac{\mathbf{s^{'}}-\min(\mathbf{s_{train}^{'}})}{\max(\mathbf{s_{train}^{'}})-\min(\mathbf{s_{train}^{'}})},\quad\mathbf{z}\in\left[0,1\right]^{d},
\end{split}
\end{equation}
where $\mathbf{z}$ is the normalized value of $d$ dimensions.

Then the input $\mathbf{z}$ is mapped to a hidden representation
$\mathbf{h}$ written as
\begin{equation}
\mathbf{h}=f_{\theta}(\mathbf{z}),
\end{equation}
where $f_{\theta}(\cdotp)$ denotes the encoder network, and $\theta$
denotes the network parameters. Since then, our framework splits into
two branches. One is for reconstruction and the other is for classification.

In the reconstruction branch, a decoder $f_{\theta^{'}}(\cdotp)$
parameterized by $\theta^{'}$ is applied to reconstruct a $\mathbf{\tilde{z}}$
from $\mathbf{h}$:
\begin{equation}
\mathbf{\tilde{z}}=f_{\theta^{'}}(\mathbf{h}),\quad\mathbf{\tilde{z}}\in\left[0,1\right]^{d}.
\end{equation}
We expect $\mathbf{\tilde{z}}$ to be as close as possible to $\mathbf{\hat{z}}$,
which is the uncorrupted input through the same preprocessing and
min-max normalization, expressed as

\begin{equation}
\mathbf{\hat{z}}=min\_max\_norm(Pre(\mathbf{s})).
\end{equation}
Each i-th training input $\mathbf{z}^{(i)}$ is mapped to a reconstruction
$\mathbf{\tilde{z}}^{(i)}$, its reconstruction target is $\mathbf{\hat{z}}^{(i)}$.
The \emph{average reconstruction error} can be evaluated by the traditional\emph{
Mean Squared Error (MSE) }as

\begin{equation}
MSE(\mathbf{\tilde{z}},\mathbf{\hat{z}})=\frac{1}{n}\sum_{i=1}^{n}\parallel\mathbf{\tilde{z}}^{(i)}-\mathbf{\hat{z}}^{(i)}\parallel^{2},
\end{equation}
where $n$ is the number of training samples.

In the classification branch, the extracted features $\mathbf{h}$
are fed into a classifier $f_{\theta_{C}}(\cdotp)$ parameterized
by $\theta_{C}$ for prediction as
\begin{equation}
\mathbf{\hat{y}}=f_{\theta_{C}}(\mathbf{h}),
\end{equation}
where $\mathbf{\hat{y}}$ is the predicted probability distribution
of all possible labels. Then the classification loss can be measured
by the\emph{ Categorical Cross Entropy }(\emph{CCE}):
\begin{equation}
CCE(\mathbf{\hat{y}},\mathbf{y})=-\frac{1}{n}\sum_{i=1}^{n}\mathbf{y}^{(i)}log(\mathbf{\hat{y}}^{(i)}),
\end{equation}
where $\mathbf{y}$ is the true label with one-hot encoding.

Parameters of this model are optimized by minimizing the whole loss:
\begin{equation}
\begin{split}\theta^{\star},\theta^{'\star},\theta_{C}^{\star}= & \arg\min_{\theta,\theta^{'},\theta_{C}}\lambda_{1}MSE(\mathbf{\tilde{z}},\mathbf{\hat{z}})+\lambda_{2}CCE(\mathbf{\hat{y}},\mathbf{y})\end{split}
,\label{eq:whole loss}
\end{equation}
where $\lambda_{1}$ and $\lambda_{2}$ represent the weight for reconstruction
loss and classification loss, respectively. This optimization will
typically be carried out by the stochastic gradient descent algorithm
and its variants.

\section{Partially Stacking Method}

\subsection{Semi-steady and steady-state RFF}

This paper aims to identify ZigBee devices for performance evaluation.
The ZigBee RF modulation format is Direct-Sequence Spread-Spectrum
(DSSS) Offset Quadrature Phase-Shift Keying (OQPSK) with half-sine
chip shaping. ZigBee signals include an eight-symbol preamble of 0x0
for each symbol. Accordingly, we will extract RFFs from these eight
preamble symbols to identify target devices.

\begin{figure}[t]
\centering\subfloat[The first three symbols.]{\includegraphics[width=1.8in]{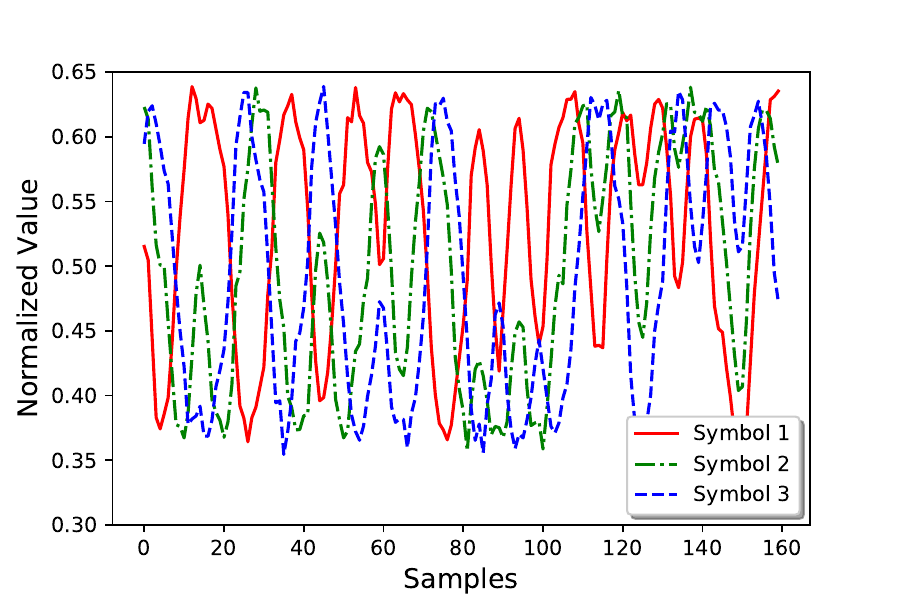}}
\subfloat[The last six symbols.]{\includegraphics[width=1.8in]{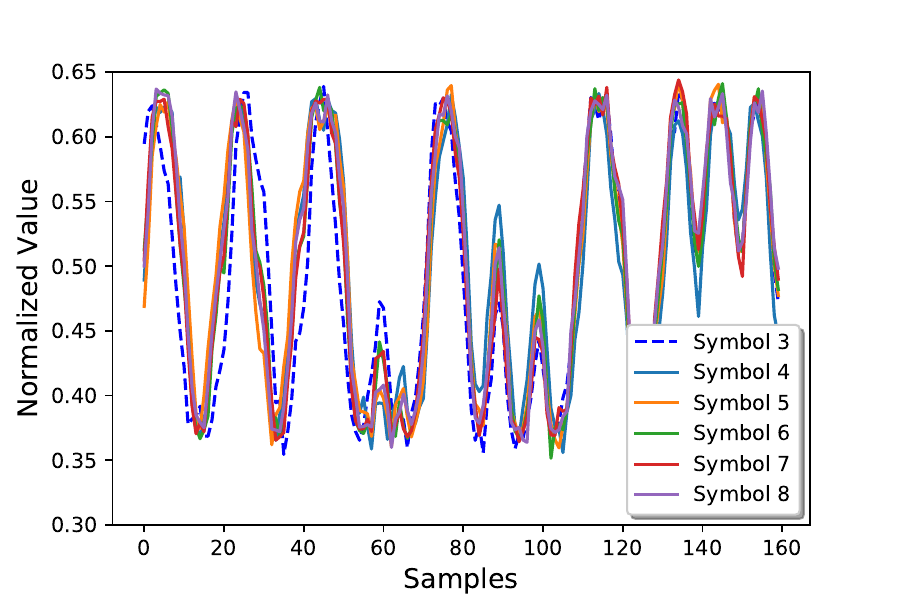}

}

\caption{The in-phase ZigBee signals of eight preprocessed preamble symbols.}
\label{Preambles Symbols}
\end{figure}

In this paper, the preprocessing module in Fig. \ref{DAE-DL-RFF}
is mainly used for synchronization, which includes timing estimation,
frequency offset compensation, and phase offset compensation. The
details can be found in \cite{Yu-2019-A}. Then we observe each preprocessed
preamble symbol to study their difference. As an example, the in-phase
signals of eight preamble symbols for one device are illustrated in
Fig. \ref{Preambles Symbols}. It is apparent that the semi-steady
portion (i.e., the first two symbols) is totally different from the
steady-state portion (i.e., the last six symbols). While the last
six symbols behave similarly. Also, the quadrature channel and other
devices exhibit similar behavior.

Thus, each preamble symbol-signal $\mathbf{s}_{k}$ can be represented
as
\begin{equation}
\mathbf{s}_{k}=RFF_{k}(\mathbf{x}_{0}),\quad k\in\left\{ 1,2,\cdots,8\right\} ,
\end{equation}
where $\mathbf{\mathbf{x}_{0}}$ denotes the identical symbol 0x0
in the preamble, and $RFF_{k}(\cdot)$ denotes the RFF function on
the k-th symbol. Since $\mathbf{s}_{1}\neq\mathbf{s}_{2}\neq\mathbf{s}_{3}$
and $\mathbf{s}_{3}\approx\mathbf{s}_{4}\approx\cdots\approx\mathbf{s}_{8}$,
it is self-evident that:
\begin{equation}
RFF_{1}(\cdot)\neq RFF_{2}(\cdot)\neq RFF_{3}(\cdot),
\end{equation}
\begin{equation}
RFF_{3}(\cdot)\approx RFF_{4}(\cdot)\approx\cdots\approx RFF_{8}(\cdot).\label{eq:Steady-state RFFs are almost the same}
\end{equation}
Therefore, if we study RFFs from the symbol scale, semi-steady RFFs
on the first two symbols are different from steady-state RFFs on the
following symbols in our target ZigBee devices.

\subsection{Partially Stacking}

Xing \emph{et al.} proposed that spread sequence in DSSS systems can
be stacked to enhance SNR for increasing identification accuracy.
However, in their simulation, RFF on each spread sequence must be
the same. Stacking is only available for signal portions with invariable
RFFs and constant data. Therefore, stacking can be applied to the
steady-state portion in the preamble rather than the semi-steady portion.
The partially stacking process can be expressed as
\begin{equation}
\begin{split}[\mathbf{s}_{1},\mathbf{s}_{2},\frac{1}{6}\sum_{i=3}^{8}\mathbf{s}_{i}]= & [RFF_{1}(\mathbf{x}_{0}),RFF_{2}(\mathbf{x}_{0}),\frac{1}{6}\sum_{i=3}^{8}RFF_{i}(\mathbf{x}_{0})]\\
= & [RFF_{1},RFF_{2},RFF_{s}](\mathbf{x}_{0})\\
= & RFF(\mathbf{x}_{0})
\end{split}
\end{equation}
where $RFF_{s}$ is the steady-state fingerprint, $RFF(\cdot)$ is
the whole fingerprint consisted of the semi-steady fingerprint and
the steady-state fingerprint.

\begin{figure}[t]
\centering

\includegraphics[width=1.7in]{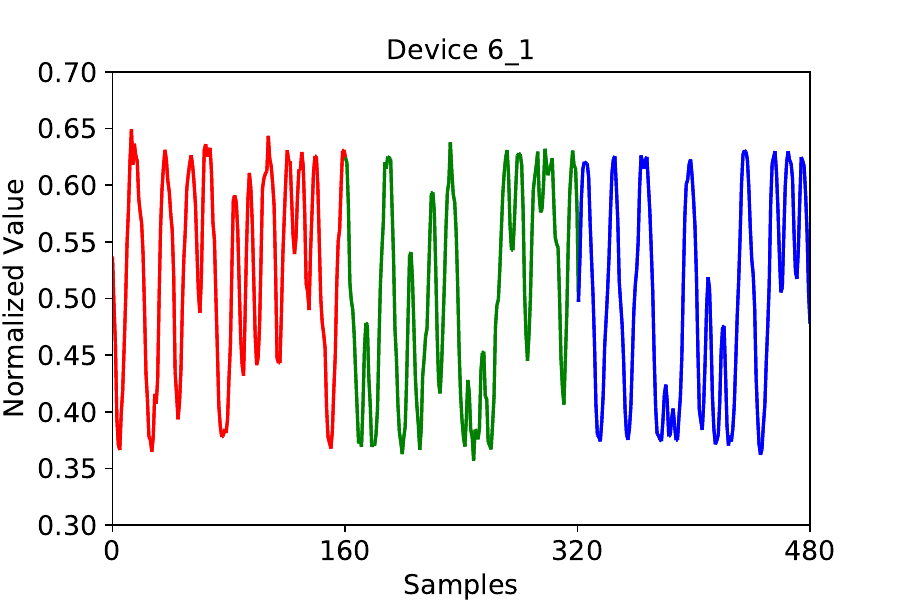}
\includegraphics[width=1.7in]{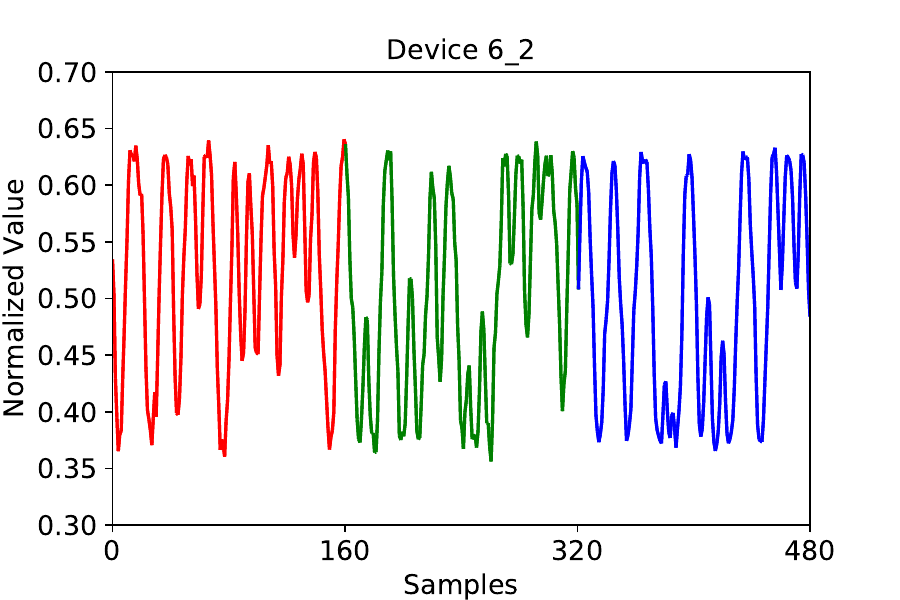}

\includegraphics[width=1.7in]{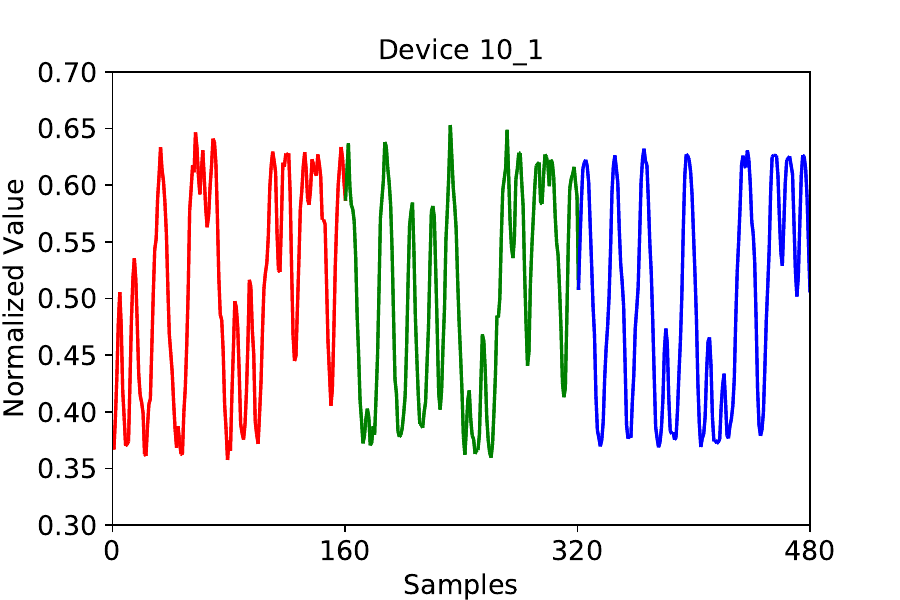}
\includegraphics[width=1.7in]{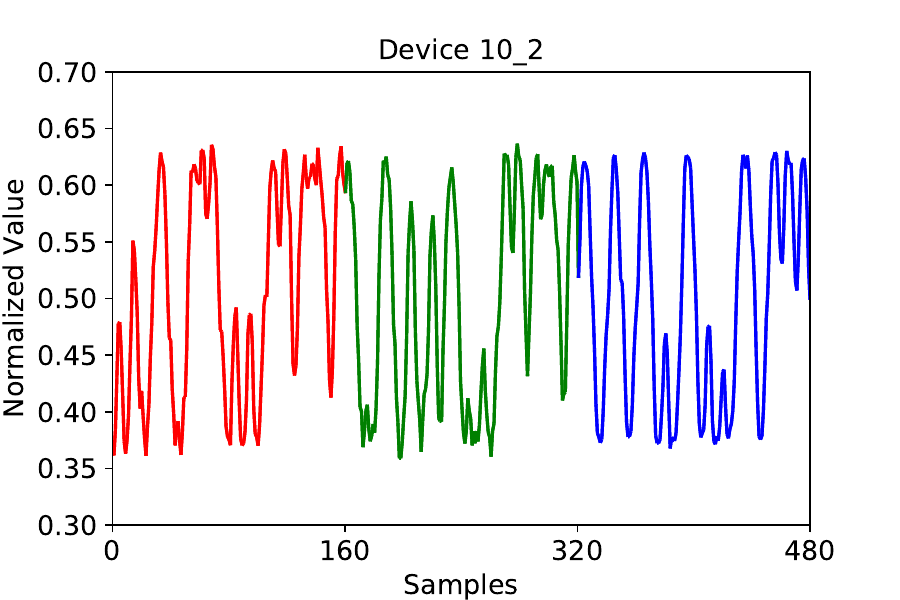}

\caption{The in-phase channels of two partially stacked preambles for two devices.}

\label{Partially stacked prambles}
\end{figure}

Partially stacking is the last step of preprocessing. After partially
stacking, two in-phase channels of two devices are randomly selected
and demonstrated in Fig. \ref{Partially stacked prambles}. It can
be seen that different devices have different RFFs, and the same device
has constant RFFs in different tests. It is also evident that the
RFFs in the first symbol differ the most, while the steady-state RFFs
in the generated last symbols by stacking are relatively similar.

\section{Results and Discussion}

\subsection{Experimental System and Data Collection\label{subsec:Experimental-System}}

Our experimental system has 27 Ti CC2530 ZigBee devices to be identified,
a USRP N210 as the receiver, and a PC with MATLAB and Tensorflow for
processing. ZigBee devices worked at their maximum power of 19 dBm,
and they were located within one meter away from the USRP, both for
enhancing the receiving SNR. In this way, the SNRs during data collection
were around 30 dB. These received signals can be regarded as approximately
ideal. Besides, owing to the 1 Mbps chip rate of CC2530, the sampling
rate of USRP was set to 10 Msps for ten-times oversampling.

In this paper, only AWGN channels were simulated as the corruption
module in Fig. \ref{DAE-DL-RFF}. AWGN channels leading to SNRs changing
from -10 dB to 30 dB with step 5 dB were all simulated. Hence, for
each simulated SNR, the corrupted input $\mathbf{z}$ of the encoder
combined with its corresponding reconstruction target $\mathbf{\hat{z}}$
and device label $\mathbf{y}$ constituted a sample $\left(\mathbf{z},\mathbf{\hat{z}},\mathbf{y}\right)$.

We had 10,944 samples, roughly 405 ones for each device. This dataset
was randomly divided into 60\% training data, 20\% validation data,
and 20\% testing data. Five-fold cross-validation was carried out
during the performance evaluation. The following networks were trained
on the dataset including all SNRs rather than a single SNR.

\subsection{Model Structure and Parameters}

Our proposed DAE-based method is a general method fit for all the
DL RFF approaches. Since CNN was mostly used in the existing DL RFFs
as stated in Section \ref{sec:Introduction}, we chose a CNN for performance
comparison. It means the encoder and the classifier constitute a CNN,
thus a CDAE comprising convolutional layers is used as the autoencoder.
The structure and parameters of our model are demonstrated in Table
\ref{Model Parameters}. The parameter selection is mainly according
to our previous work \cite{Yu-2019-A}. As described below, different
input lengths are used for comparison. When the input size is $1280\times2$
or $960\times2$, which means eight or six symbols, the max pooling
and upsampling sizes are both $4\times1$. When the input size decreases
to $480\times2$ or $320\times2$, which means three or two symbols,
the max pooling and upsampling sizes both decline to $2\times1$.
The reconstruction loss weight $\lambda_{1}$ and classification loss
$\lambda_{2}$ are set to 1 and 10, respectively. When $\lambda_{1}=0$,
our proposed CDAE model degenerates into CNN.

All our network models were trained and tested running on TensorFlow
1.12.0 with an NVIDIA GeForce GTX 1050 Ti GPU. The training was carried
out by minimizing the whole loss function in \eqref{eq:whole loss}
using an Adam solver with a batch size of 64. In addition, dropout
of 0.5 and L2 regularization of 0.001 on the dense layer were used
to prevent overfitting. The initial learning rate was set to 0.001.
The training was repeated until the validation accuracy didn't improve
within ten epochs. Then the best validation parameters were stored
for performance evaluation.

\begin{table}[t]
\begin{centering}
\centering{\small{}}
\global\long\def\arraystretch{1.2}%
{\small{} \caption{The layers and activation functions of the proposed CDAE model.}
\label{Model Parameters}}{\small\par}
\par\end{centering}
\centering{}%
\begin{tabular}{|c|c|c|c|}
\hline
\textbf{\small{}Network} & \textbf{\small{}Layer} & \textbf{\small{}Dimension} & \textbf{\small{}Activation}\tabularnewline
\hline
\multirow{1}{*}{{\small{}Input}} & \multirow{1}{*}{{\small{}Input}} & \multirow{1}{*}{1280 (or 480) $\times$ 2} & \multirow{1}{*}{{\small{}-}}\tabularnewline
\hline
\multirow{4}{*}{Encoder} & {\small{}Convolution} & 128 $\times$ (10 $\times$ 1) & {\small{}ReLU}\tabularnewline
\cline{2-4} \cline{3-4} \cline{4-4}
 & {\small{}Max Pooling} & 4 (or 2) $\times$ 1 & {\small{}-}\tabularnewline
\cline{2-4} \cline{3-4} \cline{4-4}
 & {\small{}Convolution} & 128 $\times$ (3 $\times$ 2) & {\small{}ReLU}\tabularnewline
\cline{2-4} \cline{3-4} \cline{4-4}
 & {\small{}Max Pooling} & 4 (or 2) $\times$ 1 & {\small{}-}\tabularnewline
\hline
\multirow{5}{*}{Decoder} & {\small{}Convolution} & 128 $\times$ (3 $\times$ 2) & {\small{}ReLU}\tabularnewline
\cline{2-4} \cline{3-4} \cline{4-4}
 & UpSampling & 4 (or 2) $\times$ 1 & {\small{}-}\tabularnewline
\cline{2-4} \cline{3-4} \cline{4-4}
 & {\small{}Convolution} & 128 $\times$ (10 $\times$ 1) & {\small{}ReLU}\tabularnewline
\cline{2-4} \cline{3-4} \cline{4-4}
 & UpSampling & 4 (or 2) $\times$ 1 & {\small{}-}\tabularnewline
\cline{2-4} \cline{3-4} \cline{4-4}
 & {\small{}Convolution} & 1 $\times$ (3 $\times$ 1) & Sigmoid\tabularnewline
\hline
\multirow{4}{*}{Classifier} & {\small{}Flatten} & - & {\small{}-}\tabularnewline
\cline{2-4} \cline{3-4} \cline{4-4}
 & {\small{}Dense} & 1024 & {\small{}ReLU}\tabularnewline
\cline{2-4} \cline{3-4} \cline{4-4}
 & {\small{}Dropout (0.5)} & - & {\small{}-}\tabularnewline
\cline{2-4} \cline{3-4} \cline{4-4}
 & {\small{}Dense} & 27 & {\small{}Softmax}\tabularnewline
\hline
\end{tabular}
\end{table}

\subsection{CNN vs. CDAE and Semi-steady RFF vs. Steady-state RFF}

\begin{table*}[t]
\caption{The identification rates using different networks and inputs.}
\label{CNN vs CDAE with various inputs}

\centering
\renewcommand{\arraystretch}{1.2}{
\centering{}%
\begin{tabular}{|c|c|c|c|c|c|c|c|c|}
\hline
SNR (dB) & CNN\_2 (\%) & CDAE\_2 (\%) & CNN\_6 (\%) & CDAE\_6 (\%) & SCNN\_6 (\%) & SCDAE\_6 (\%) & CNN\_8 (\%) & CDAE\_8 (\%)\tabularnewline
\hline
-10 & \textit{27.3} & \textbf{28.7} & 12.4 & 14.4 & 15.4 & 17.2 & 23.3 & 23.6\tabularnewline
\hline
-5 & \textit{50.7} & \textbf{52.0} & 24.8 & 26.9 & 31.0 & 35.9 & 43.5 & 47.3\tabularnewline
\hline
0 & \textit{63.4} & \textbf{66.1} & 32.5 & 34.0 & 33.5 & 42.7 & 52.6 & 58.3\tabularnewline
\hline
5 & \textit{81.9} & \textbf{84.3} & 48.1 & 53.6 & 49.2 & 58.3 & 77.3 & 82.0\tabularnewline
\hline
10 & 90.7 & 91.4 & 74.8 & 80.4 & 72.1 & 81.1 & \textit{90.7} & \textbf{94.6}\tabularnewline
\hline
15 & 94.6 & 95.5 & 90.9 & 91.7 & 83.5 & 93.0 & \textit{96.8} & \textbf{98.2}\tabularnewline
\hline
20 & 96.8 & 97.1 & 94.5 & 94.9 & 87.3 & 96.7 & \textit{98.1} & \textbf{99.4}\tabularnewline
\hline
25 & 96.4 & 96.2 & 96.5 & 97.2 & 90.3 & 98.2 & \textit{97.4} & \textbf{99.6}\tabularnewline
\hline
30 & 96.9 & 97.7 & 97.4 & 98.4 & 92.6 & 98.5 & \textit{98.0} & \textbf{99.8}\tabularnewline
\hline
\end{tabular}}
\end{table*}
As shown in Table \ref{CNN vs CDAE with various inputs}, we conducted
experiments both for CNNs and CDAEs using different signal portions
as inputs. Obviously, our proposed CDAE models achieve better performance
at all scenarios compared to CNNs. The main performance improvement
occurs in the region of {[}-5,10{]} dB.

Two-symbol semi-steady portions are fed into CNN\_2 and CDAE\_2 to
extract the semi-steady RFF. It can be seen that the semi-steady RFF
performs the best at low SNRs from -10 dB to 5 dB compared to all
other inputs. The semi-steady RFF is relatively robust to AWGN channels.

The inputs of CNN\_6 and CDAE\_6 are six-symbol steady-state portions.
Compared to CNN\_2 and CDAE\_2, they only behave better at very high
SNRs, which means that the steady-state RFF has more fine-grained
features. However, by stacking these six symbols, SCDAE\_6 outperforms
CDAE\_6 at all SNRs, while SCNN\_6 is worse than CNN\_6 when SNR is
not lower than 0 dB. It can be seen that CDAE can recover these fine-grained
features from stacking as opposed to CNN. Hence, only stacking with
appropriate processing can improve the identification rate. In a word,
the semi-steady RFF contributes more than the steady-state RFF in
ZigBee device identification.

Besides, we aim to evaluate the performance by using these two RFFs
simultaneously. Therefore, the whole preamble is sent into CNN\_8
and CAE\_8 for identification, which has been employed in \cite{Merchant-2018-Deep}.
In this way, the identification accuracy improves at high SNRs ($\geq10dB$),
while its performance at low SNRs is still a bit poor in contrast
to using the semi-steady RFF only.

\subsection{Partially Stacking and Final Performance}

\begin{table*}[tp]
\caption{Accuracies, precisions, recalls and their 95\% confidence intervals
for partially stacking-based CNN and CDAE models.}
\label{Final performance}

\centering
\renewcommand{\arraystretch}{1.2}{
\centering{}%
\begin{tabular}{|c|c|c|c|c|c|c|}
\hline
 & \multicolumn{3}{c|}{PSCNN} & \multicolumn{3}{c|}{PSCDAE}\tabularnewline
\hline
SNR (dB) & Accuracy (\%) & Precision (\%) & Recall (\%) & Accuracy (\%) & Precision (\%) & Recall (\%)\tabularnewline
\hline
-10 & $36.3\pm2.1$ & $34.7\pm3.0$ & $34.7\pm2.0$ & \textbf{38.7$\pm$0.6} & $39.4\pm0.5$ & $37.0\pm0.7$\tabularnewline
\hline
-5 & $59.4\pm2.0$ & $61.2\pm2.2$ & $59.0\pm2.3$ & \textbf{63.4$\pm$1.3} & $65.7\pm1.3$ & $63.0\pm1.1$\tabularnewline
\hline
0 & $71.5\pm1.9$ & $72.1\pm2.0$ & $70.5\pm2.3$ & \textbf{76.1$\pm$1.8} & $78.0\pm1.8$ & $76.3\pm2.0$\tabularnewline
\hline
5 & $86.3\pm2.3$ & $87.8\pm2.1$ & $85.8\pm3.0$ & \textbf{91.3$\pm$0.9} & $92.4\pm1.0$ & $91.2\pm1.0$\tabularnewline
\hline
10 & $95.7\pm1.9$ & $96.5\pm1.8$ & $95.6\pm2.6$ & \textbf{97.5$\pm$0.5} & $97.9\pm0.5$ & $97.7\pm0.5$\tabularnewline
\hline
15 & $97.0\pm1.1$ & $97.3\pm0.8$ & $96.4\pm1.1$ & \textbf{98.9$\pm$0.6} & $99.2\pm0.4$ & $99.0\pm0.6$\tabularnewline
\hline
20 & $98.6\pm0.9$ & $98.8\pm0.6$ & $98.3\pm1.0$ & \textbf{99.7$\pm$0.3} & $99.7\pm0.3$ & $99.7\pm0.3$\tabularnewline
\hline
25 & $98.7\pm0.9$ & $98.5\pm0.6$ & $98.0\pm0.9$ & \textbf{99.1$\pm$0.5} & $99.4\pm0.4$ & $99.3\pm0.5$\tabularnewline
\hline
30 & $99.0\pm0.9$ & $98.7\pm0.9$ & $98.5\pm1.1$ & \textbf{99.0$\pm$0.5} & $99.3\pm0.4$ & $99.2\pm0.4$\tabularnewline
\hline
\end{tabular}}
\end{table*}
\begin{figure}[t]
\includegraphics[width=3.5in]{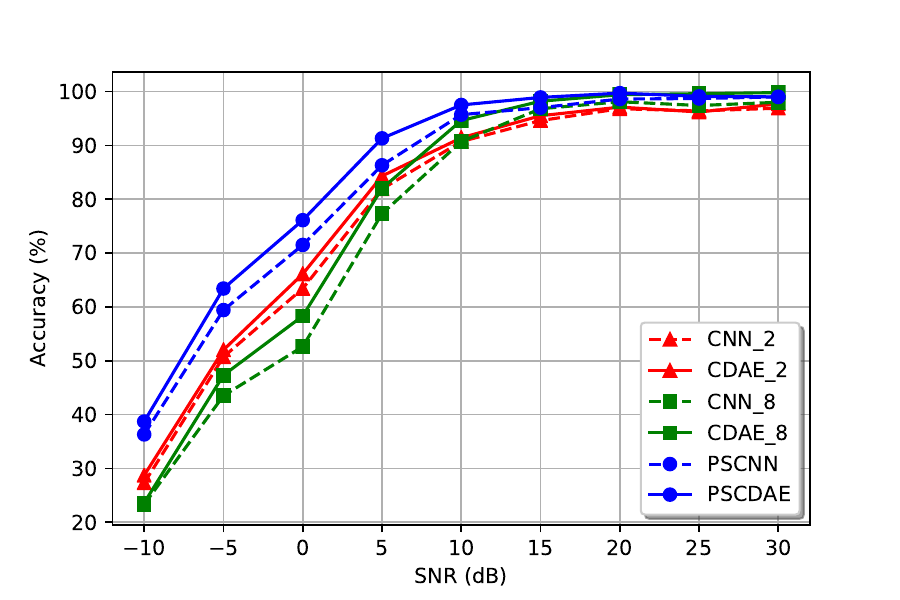}\caption{Performance comparison of CDAE and CNN by combining semi-steady and
steady-state RFFs differently.}
\label{Performance Comparison}
\end{figure}
Since stacking with CDAE (SCDAE\_6) works better than CDAE without
stacking (CDAE\_6), it is reasonable to imagine that PSCDAE can further
improve the identification accuracy. The identification accuracies,
precisions, recalls and their 95\% confidence intervals of PSCDAE
and Partially Stacking-based CNN (PSCNN) are manifested in Table \ref{Final performance}.

Besides, for easy comparison with SCDAE\_2 and SCDAE\_8 which behave
the best in different SNR regions in Table \ref{CNN vs CDAE with various inputs},
the classification accuracies of six approaches at all SNRs are illustrated
in Fig. \ref{Performance Comparison}. The dash lines represent CNN-based
methods, and the solid lines represent CDAE-based methods. Besides,
lines with the same marker have the same input. It is apparent that
CDAE-based methods behave better than corresponding CNN-based methods,
especially at SNRs from -5 dB to 10 dB. Furthermore, PSCDAE has the
best accuracy at almost all SNRs, which demonstrates that it can combine
semi-steady and steady-state RFFs effectively. It is also noticeable
that PSCDAE is a bit worse than CDAE\_8 by $5\permil$ and $8\permil$
at 25 dB and 30 dB, respectively. This is owing to the information
loss during stacking. As stated in \eqref{eq:Steady-state RFFs are almost the same},
the steady-state RFFs in the six preamble symbols are almost equal,
but some different fine-grained features still exist. These fine-grained
features can be extracted by CDAE\_8 at high SNRs for identification.
While as SNR goes down, these features will be gradually covered by
noise.

By comparison with CNN\_8 which is used in \cite{Merchant-2018-Deep},
PSCDAE can respectively improve the identification accuracy by 15.4\%,
19.9\%, 23.5\%, 14.0\%, 6.8\%, 2.1\%, 1.6\%, 1.7\%, 1.0\% at SNRs
from -10 dB to 30 dB with step 5 dB. Besides, according to Table \ref{Final performance},
it is also crystal-clear that the 95\% confidence intervals for PSCDAE
are almost only half of that for PSCNN. This is to say, models trained
by PSCDAE are more precise than those trained by PSCNN. In other words,
performance fluctuations of PSCDAE are smaller. Furthermore, it can
be seen that the accuracy of PSCDAE first increases dramatically and
then slightly goes down, it reaches the top when SNR is equal to 20
dB. This is because, at high SNRs, the gain introduced by the autoencoder
is gradually lost. Due to the existing reconstruction loss in the
optimization objective, those fine-grained features that are useful
for classification while adverse to reconstruction cannot be learned
by PSCDAE.

\section{Conclusions}

This paper has demonstrated a universal DAE-based framework for DL
RFF approaches. A partially stacking technique has also been proposed
to leverage both the semi-steady and steady-state RFFs efficiently
for identifying ZigBee devices. A two-layer CNN and the corresponding
CDAE model have been used to show the superiority of our proposed
scheme. We have conducted various experiments on our testbed which
includes a USRP as the receiver and 27 CC2530 nodes as targets. Experimental
results demonstrate that our proposed PSCDAE outperforms the traditional
CNN in \cite{Merchant-2018-Deep} by 14\% to 23.5\% at low SNRs (-10
dB to 5 dB) under AWGN channels in terms of identification accuracy.
At high SNRs, it can also slightly improve performance. Besides, the
models trained by PSCDAE are much more robust than those trained by
CNN. In future work, we will focus on studying multipath channels
as the corruption module as well as using images derived from transforms
as the input.

\appendices{}

\section*{Acknowlegment}

This work was supported in part by the National Natural Science Foundation
of China under Grant 61571110, 61602113, 61601114, 61801115, Purple
Mountain Laboratories (PML) and Campus France.

\small\bibliographystyle{IEEEtran}
\bibliography{mybib}

\end{document}